\documentclass{Interspeech}

\interspeechcameraready 

\usepackage{tabularx}

\title{Low-resource keyword spotting using contrastively trained transformer acoustic word embeddings}

\author{Julian}{Herreilers}
\author{Christiaan}{Jacobs}
\author{Thomas}{Niesler}

\affiliation{Department of Electrical and Electronic Engineering}{Stellenbosch University}{South Africa}
\email{jherreilers@gmail.com, christiaanjacobs97@gmail.com, trn@sun.ac.za}

\usepackage{comment}

\begin{document}

\newcommand{\trn}[1]{\textcolor{red}{[trn: #1]}} 

\newcommand{\jth}[1]{\textcolor{blue}{[jth: #1]}} 

\newcommand{\cj}[1]{\textcolor{orange}{[cj: #1]}} 

\maketitle

\begin{abstract}
    
We introduce a new approach, the ContrastiveTransformer, that produces acoustic word embeddings (AWEs) for the purpose of very low-resource keyword spotting.
The ContrastiveTransformer, an encoder-only model, directly optimises the embedding space using normalised temperature-scaled cross entropy (NT-Xent) loss.
We use this model to perform keyword spotting for radio broadcasts in Luganda and Bambara, the latter a severely under-resourced language.
We compare our model to various existing AWE approaches, including those constructed from large pre-trained self-supervised models, a recurrent encoder which previously used the NT-Xent loss, and a DTW baseline.
We demonstrate that the proposed contrastive transformer approach offers performance improvements over all considered existing approaches to very low-resource keyword spotting in both languages.
\end{abstract}

\section{Introduction}

The United Nations have for a number of years funded the development of keyword spotting (KWS) systems that can be applied to radio broadcasts as a way for the organisation to monitor and inform its humanitarian relief efforts~\cite{unpulse_whie_paper,LR_KWS_journal}.
These systems operate in areas of the developing world with little or no internet infrastructure, where communities must rely on community radio stations to relay concerns, opinions and information.
Usually, the languages concerned are severely under-resourced and the urgency of the humanitarian crisis requires systems to be put in place very quickly. 
These constraints make KWS that relies on automatic speech recognition (ASR) followed by a text-based search infeasible.  

An alternative approach to KWS is query-by-example (QbE), which requires as little as a single spoken keyword template to search an unlabelled audio corpus.
This is extremely appealing in this resource and time-constrained scenario, because such templates are much easier to acquire in a new language. 
Dynamic Time Warping (DTW) is a well-established choice for matching such query templates to search audio~\cite{SegmentalDTW_2009}. 
However, DTW is computationally expensive. 
In previous work, we improved retrieval speed by using DTW scores to train a convolutional neural network, which is computationally more efficient~\cite{menon2018fastasrfree_cnndtw_intro,menon2018asrfree_bnf_intro,menon2019feature_cAE_AE_intro}.
This was further improved using human feedback to correct certain KWS errors \cite{KWS_HITL}.

Acoustic word embeddings (AWEs), which map variable-length sequences to fixed-dimensional vectors, are an appealing recent approach to QbE. 
AWEs eliminate the need for the repeated alignments of DTW by comparing a query vector with a vector representation of the segment to be searched.
This has been demonstrated to match or improve on the performance of DTW while being more computationally efficient \cite{Levin2013FixeddimensionalAE_laplacian_eignemaps, Settle_multilingual_QBE_search}. 
AWEs can be realised using various neural network architectures, notably RNNs \cite{Chung2016AudioWU_AERNN, Kamper2018TrulyUA_CAE_RNN, Settle2016DiscriminativeAW_usingRNN, Staden2020ACO_CPC_with_CAERNN,jacobs2021acousticwordembeddingszeroresource_contrastivernn} and transformers \cite{Lin2023SelfSupervisedAW_CTE}. 
For example, an effective approach for low-resource settings has been to train a multilingual AWE model on well-resourced languages, and then apply it without adjustment to the target low-resource language~\cite{jacobs2021acousticwordembeddingszeroresource_contrastivernn,kamper2021improved_AWE_multilingual, jacobs2021multilingualtransferacousticword_related_lang}.

More recently, AWEs have made direct use of the features calculated by large transformer-based models, which have themselves been trained using self-supervised learning (SSL) on extremely large datasets \cite{sanabria2023analyzing_embeddings_pretrained_models, sanabria2023acoustic_continued_pretraining}. 
This is a particularly promising approach in highly resource-constrained environments where there is insufficient data for fine-tuning on the target language.
{In~\cite{jacobs2023_hs_detection}, a multilingual AWE model outperformed an ASR system in a hate-speech detection task on out-of-domain radio broadcasts, even though the ASR model had been fine-tuned on one hour of data from the target language.}

In this work we use AWEs to perform KWS on radio broadcasts in three languages: English, our development language, Luganda (Ganda), a low-resource Bantu language spoken in Uganda, and Bambara, a very low-resource Mande language spoken in Mali. 
To achieve this, we propose a new approach, the ContrastiveTransformer, which applies contrastive loss to a transformer encoder~\cite{vaswani2023attention} to train an AWE model that is subsequently applied to KWS on radio broadcasts.
This extends previous work proposing transformer encoders for AWEs~\cite{Lin2023SelfSupervisedAW_CTE} and applying a contrastive loss to RNNs~\cite{jacobs2021acousticwordembeddingszeroresource_contrastivernn}. 
We show how this outperforms other recent and established approaches when performing KWS on our highly under-resourced languages.

We make the following contributions towards improving severely under-resourced keyword spotting. (1) We are the first to compare a variety of different embedding techniques to KWS applied to radio broadcasts. (2) We demonstrate that using contrastively-trained AWEs for KWS offers consistent improvements over a previous reconstruction-based approach to this task. (3) We propose a new contrastive transformer-based model which achieves improved performance in our low-resource setting while requiring only a handful of isolated keyword templates to operate.

\section{Acoustic word embedding (AWE) techniques}
\label{sec:AWE_techniques}
We evaluate and compare five AWE modelling approaches for QbE using radio speech. 
One of these five is the ContrastiveTransformer we propose.
The other four are (i)~a correspondence autoencoder RNN (CAE-RNN) as proposed in~\cite{Kamper2018TrulyUA_CAE_RNN} and applied to hate speech detection in~\cite{jacobs2023_hs_detection}, (ii)~meanpooling and (iii)~subsampling of self-supervised features, as proposed in~\cite{sanabria2023analyzing_embeddings_pretrained_models}, and (iv)~the ContrastiveRNN, as proposed in~\cite{jacobs2021acousticwordembeddingszeroresource_contrastivernn}.


\subsection{CAE-RNN}
\label{sec:cae_rnn}
For the CAE-RNN, an encoder RNN inputs a feature sequence $\boldsymbol{X}$ corresponding to a single word
type.
The final encoder hidden state is then passed as input to a decoder RNN, which is trained to reconstruct a feature sequence $\boldsymbol{X}'$ corresponding to a different instance of the
input
word type. 
After training, the final encoder hidden state $\boldsymbol{w}$ is used as the AWE. 
Training proceeds by minimising the reconstruction loss between $\boldsymbol{X}'$ and the decoder output sequence~\cite{Kamper2018TrulyUA_CAE_RNN}.

\subsection{Meanpooling and subsampling of self-supervised features}
\label{sec:meanpool_and_subsample_techniques}

Meanpooling constructs an AWE by averaging all the features extracted  from a segment to produce a single embedding~\cite{sanabria2023analyzing_embeddings_pretrained_models}.
Subsampling concatenates $K$ equally-spaced feature vectors sampled from $\boldsymbol{X}$ to produce the AWE~\cite{Levin2013FixeddimensionalAE_laplacian_eignemaps, sanabria2023analyzing_embeddings_pretrained_models}.

\subsection{ContrastiveRNN}
\label{sec:contrastive_rnn}
This model consists of only an RNN encoder which is optimised directly in the embedding space by contrastive learning~\cite{jacobs2021acousticwordembeddingszeroresource_contrastivernn}. 
Consider a batch consisting of $N$ distinct positive pairs $(\boldsymbol{X}^{(a,i)}, \boldsymbol{X}^{(p,i)})$, where $i = 1,2\ldots N$.
Each pair consists of an anchor $\boldsymbol{X}^{(a,i)}$ and a second sequence of the same type $\boldsymbol{X}^{(p,i)}$, with corresponding embeddings $\boldsymbol{w}^{(a,i)}$ and $\boldsymbol{w}^{(p,i)}$.
The NT-Xent loss minimises the distance in the embedding space between the two sequences in a positive pair while maximising the distance between the anchor $\boldsymbol{w}^{(a,i)}$ of this positive pair and all $2N-2$ sequences in the other pairs:

\begin{equation}
\mathcal{L}_i = 
-\log
\frac{\text{exp$(\text{sim}(\boldsymbol{w}^{(a,i)}, \boldsymbol{w}^{(p,i)})/\tau$)}}
     {\sum\limits_{\forall \boldsymbol{w} \in \mathcal{W} }\text{exp$(\text{sim}(\boldsymbol{w}^{(a,i)}, \boldsymbol{w})/\tau)$}
     }
\label{eq:contrastive_RNN_loss}
\end{equation} 
where $\mathcal{W} = \left\{\boldsymbol{w}^{(a,j)} \cup \boldsymbol{w}^{(p,k)} \right\} $ and $j \in \{1,2,\ldots N\}$ with the exclusion of $j=i$ and $k \in \{1,2,\ldots N\}$.
$\tau$ and 
$\text{sim}(\cdot)$
are the temperature coefficient and cosine similarity respectively. 
This loss is summed over all pairs in the batch.




\subsection{ContrastiveTransformer}
\label{sec:contrastive_transformer}

\begin{figure}[!t]
    \centering
    \includegraphics[width=1\linewidth]{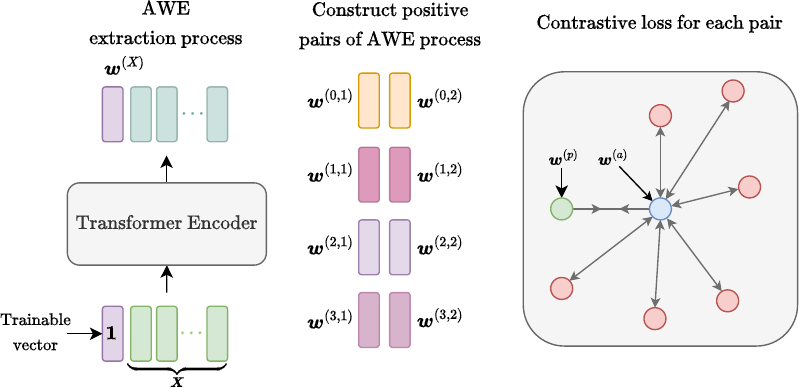}
    \caption{ContrastiveTransformer architecture. A trainable vector is prepended to the input sequence and processed by the transformer. The first output vector of the final layer is used as the AWE $\boldsymbol{w}$. 
    The model is trained using a batch of distinct pairs, where each pair consists of two different examples of the same word type. During training, the distance between two embeddings of the same word type is minimised while the distance between embeddings of differing word types is maximised.}
    \label{fig:contrastive_transformer_architecture}
\end{figure}

Taking inspiration from \cite{Lin2023SelfSupervisedAW_CTE} which proposes a correspondence-styled teacher-student AWE transformer, we consider using only a transformer encoder 
which we train using the loss function in
Equation~\ref{eq:contrastive_RNN_loss}.
In order to obtain a fixed-dimensional representation for an input sequence $\boldsymbol{X}$, we prepend a trainable vector 
of ones
before encoding the sequence~\cite{Lin2023SelfSupervisedAW_CTE,devlin2019bert}. 
Thereafter, we use the first vector produced at the final transformer layer as the
AWE
$\boldsymbol{w}$.
This procedure is illustrated in Fig.~\ref{fig:contrastive_transformer_architecture}. 

\section{Keyword spotting using AWEs}

When performing keyword spotting (KWS)
by
QbE, a collection of isolated keyword
templates is used
 to query whether the associated keyword is present in an unlabelled utterance~\cite{jacobs2023_hs_detection}. 
Since word boundaries are not available for the search utterance, a variable-length window can be applied to divide the utterance into overlapping segments, each of which is then subsequently embedded. 
The cosine distances between the embeddings of the keyword template and the search segments are calculated as a measure of similarity.
Finally, a threshold is applied to make a decision on a keyword's presence~\cite{Settle_multilingual_QBE_search}.

\begin{figure*}[!t]
  \includegraphics[width=\textwidth]{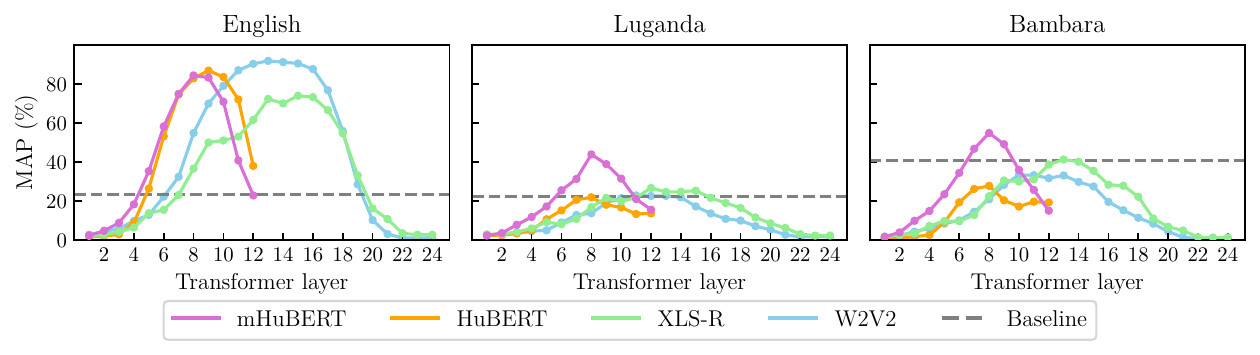}
  \caption{KWS results on the development set. AWEs are obtained through meanpooling features extracted from different transformer layers. The baseline corresponds to DTW using bottleneck features~\cite{KWS_HITL}.}
  \label{fig:meanpooling_results}
\end{figure*}

\section{Datasets}
\label{sec:Dataset_section}
Different datasets are used for AWE training and for subsequent KWS.
\subsection{AWE training}
For the supervised training of multilingual AWE models, as required for methods 2.1, 2.3 and 2.4 in Section~\ref{sec:AWE_techniques}, we utilise the NCHLT dataset \cite{NCHLT_paper}.
This corpus contains the 11 official spoken languages of South Africa, 9 of which belong to the Southern Bantu family and are distantly related to Luganda, a Northeast Savannah Bantu language. 
In addition, we include Swahili, a fairly well-resourced central African Bantu language in the same family as Luganda, by extracting a subset of the Swahili data included in the Common Voice \cite{Ardila2019CommonVoice} corpus that is of a similar size to each NCHLT language. 
This follows ~\cite{jacobs2021multilingualtransferacousticword_related_lang}, which shows that training AWE models with languages related to the target low-resource language improves downstream performance.

\subsection{Keyword spotting}
We perform KWS for three languages: English, which acts as our development language for hyperparameter selection, and two under-resourced evaluation languages, Luganda and Bambara. 
For each language, we use two corpora for KWS~\cite{menon2019feature_cAE_AE_intro,KWS_HITL}.
The first is a collection of isolated keyword templates and is the only labelled data used for our KWS system.
The second is the search corpus, and consists of audio collected from local radio stations, split into a development set for hyperparameter tuning and a test set for final evaluation.
These two corpora are summarised in Tables~\ref{tbl:kws_template_corpus} and~\ref{tbl:KWS_radio_search_corpus}. {The search corpus for our evaluation languages (Luganda and Bambara) are noisier than our development language (English) due to transmission distortion present in the broadcasts \cite{KWS_HITL}. 

\begin{table}[!h]
    \renewcommand{\arraystretch}{1.1}
    \centering
    \caption{Corpus of isolated keyword templates.}
    \begin{tabular}{@{}ccccc@{}}
        \toprule
        Language & Speakers & Keyword & Total & Duration \\
         &  & types & utterances & \\
        \midrule
        English & $24$ & $40$ & $1920$ & $34.27$m \\
        Luganda & $16$ & $18$ & $603$ & $13.83$m \\
        Bambara & $139$ & $30$ & $8335$ & $2.05$h \\
        \bottomrule
    \end{tabular}
    \label{tbl:kws_template_corpus}
\end{table}

        

\begin{table}[!h]
    \renewcommand{\arraystretch}{1.1}
    \centering
    \caption{Search corpus of radio speech.}
    \begin{tabular}{@{}lcccc@{}}
        \toprule
        & \multicolumn{2}{c}{Development set} & \multicolumn{2}{c}{Test set } \\
           & Utterances & Duration & Utterances & Duration\\
        \midrule
        English & $4220$ & $7.80$h & $5005$ & $10.28$h \\
        Luganda & $6052$ & $5.57$h & $1420$ & $1.99$h \\
        Bambara & $11715$ & $7.45$h & $12790$ & $7.77$h \\
        \bottomrule
    \end{tabular}
    \label{tbl:KWS_radio_search_corpus}
\end{table}

\section{Experimental setup}
\label{sec:Experimental_setup}

\subsection{Pre-trained models}
\label{sec:SSL_models_experiments}

For meanpooling and subsampling as described in Section~\ref{sec:meanpool_and_subsample_techniques},
we consider four pre-trained self-supervised models from the wav2vec2.0 and HuBERT families~\cite{baevski2020wav2vec2,HuBERT_hsu2021}.
For the former, we consider the English wav2vec2.0 LARGE (W2V2) and the multilingual XLS-R models~\cite{XLS-R_babu2021}. 
For the latter, we consider the newly-released mHuBERT-147 (MHB) which has been trained on more languages than XLS-R and shows competitive performance while being of the smaller BASE configuration~\cite{ZanonBoito2024mHuBERT147AC}.
We also include the monolingual (English) HuBERT BASE (HB) to evaluate the effect of multilingual training.
Note that Luganda is part of the training sets for MHB and XLS-R, while Bambara forms part of neither.


\begin{table*}[!t]
    \centering
    \scriptsize
    \caption{Development set keyword spotting results achieved with various acoustic word embeddings (AWEs).}
    \begin{tabularx}{.9\linewidth}{@{\extracolsep{4pt}}lcccccccccc}
        \toprule
            &&
            \multicolumn{3}{c}{English}  & 
            \multicolumn{3}{c}{Luganda}  & 
            \multicolumn{3}{c}{Bambara}  \\
          \cmidrule(lr){3-5} \cmidrule(lr){6-8} \cmidrule(lr){9-11}
          Architecture&Configuration& MAP & P@10 & P@N & MAP & P@10 & P@N & MAP & P@10 & P@N \\ 
          \midrule
        DTW with BNF &(\textbf{Baseline} \cite{KWS_HITL})& 23.37 & 41.11 &25.96 & 22.44 & 52.50 & 25.92 & 40.90 & 56.09 & 40.48 \\
        
        mHuBERT-147 & Meanpooling (layer 8) & 84.34 & 91.67 & 78.97 & 43.87 & 72.50 & 45.11 & 54.85 & 73.91 & 53.22\\

         mHuBERT-147 & Subsampling (layer 8) & 79.77 & 85.28 & 74.87 & 32.45 & 60.00 & 35.80 & 44.43 & 58.70 & 45.79\\
        

        CAE-RNN & afr+sw+xho+sot & 78.56 & 89.72 & 74.57 & 47.19 & 70.00 & 48.47 & 62.53 & 76.09 & 60.98\\

        ContrastiveRNN&afr+sw+xho+sot & 86.41 & 92.78 & 82.24 & 65.74 & \textbf{89.17} & 63.48 & 70.96 & 80.87 & 69.08\\

        ContrastiveTransformer & afr+sw+xho+sot & \textbf{89.16} & \textbf{95.00} & \textbf{85.67} & \textbf{69.91} & 87.50 & \textbf{68.81} & \textbf{74.53} & \textbf{83.91} & \textbf{73.12} \\

        \bottomrule
    \end{tabularx}
    \label{tbl:multilingual_AWE_dev_results}
    \vspace*{-3mm}
\end{table*}

\subsection{AWE models and input features}
\label{sec:model_sizes}
We train the multilingual CAE-RNN, ContrastiveRNN and ContrastiveTransformer models in a supervised manner using word pairs extracted from the NCHLT corpus, following the procedure described in~\cite{jacobs2021multilingualtransferacousticword_related_lang}.
Word-level
alignments for Swahili were obtained using the Montreal Forced Aligner~\cite{McAuliffe2017MontrealFA}.
Alignments for the other languages were already available.

For CAE-RNN and ContrastiveRNN we use the same architecture proposed in \cite{jacobs2021acousticwordembeddingszeroresource_contrastivernn,jacobs2021multilingualtransferacousticword_related_lang}. 
Our ContrastiveTransformer is a 3-layer transformer encoder, each with 16 attention heads, followed by a linear layer to produce a 256-dimensional AWE. 
This architecture was determined by optimisation on our development sets.

We considered various language configurations for AWE training, always excluding our evaluation languages. 
Bantu languages were added in an effort to improve performance for Luganda. 
Bambara belongs to the Mande language family, for which we have no related languages.
Here we hypothesised that including a variety of different languages would be beneficial. 
We chose Xitsonga (tso) and Tshivenda (ven) as development languages for training our AWEs using the \textit{same-different task}, which evaluates the model's ability to generate similar AWEs for the same words uttered by different speakers~\cite{carlin11_interspeech}. 
To allow for a fair comparison, each AWE model was trained on the same 400k word pairs drawn from the languages Afrikaans (afr), Swahili (sw), isiXhosa (xho) and Sesotho (sot).
During training, both $(\boldsymbol{X}, \boldsymbol{X}')$ and $(\boldsymbol{X}', \boldsymbol{X})$ pairs were used. 

In preliminary experiments, by comparing different input features, we found that using the outputs of the 10th transformer layer of MHB achieves the same performance for unseen languages as the 13th layer of XLS-R~\cite{jacobs2023_hs_detection},  while offering improved performance for seen languages (Swahili).

\subsection{QbE system and evaluation metrics}
\label{sec:QBE_system}
 
AWEs were extracted for all keyword templates (Section~\ref{sec:AWE_techniques}). 
Subsampled AWEs were obtained by concatenating 10 equally-spaced vectors, as recommended in~\cite{sanabria2023analyzing_embeddings_pretrained_models},
resulting in a 7680-dimensional AWE for MHB. 
 
To achieve KWS, a search window with a length varying between 10 and 65 frames was swept over the search sequence, using a stride of 5. 
As was also noted in~\cite{sanabria2023analyzing_embeddings_pretrained_models}, normalisation prior to AWE extraction leads to substantial performance improvements.
We apply mean and variance normalisation on a per-speaker basis for the keyword templates and on a per-utterance basis for the search utterances. 

 
By applying a threshold to all similarity scores, we compute a per-keyword average precision (AP). 
The precision at 10 (P@10) and P@N are determined for each keyword to monitor the system's retrieval performance. 
These two
metrics
indicate the percentage of the top 10 or top N retrieved utterances that in fact contain the keyword, where N is the total number of utterances in which the keyword truly appears.
We report the mean average precision (MAP), mean P@10 and mean P@N across all keywords. 
To ensure that P@10 values are meaningful, we only report results for keywords which appear in at least 10 utterances in both the development and test sets.
This leaves 36, 12 and 23 keyword types for English, Luganda and Bambara respectively. 
As a baseline, we compare performance to DTW with bottleneck features (BNF) as this is reported in~\cite{KWS_HITL} for Bambara and English.




\section{Results}
\label{sec:Results}





\subsection{Keyword spotting by meanpooling of self-supervised features}
\label{sec:Layerwise_SSL_Meanpooling_results}

A layer-wise analysis of self-supervised model features, similar to that performed in \cite{sanabria2023analyzing_embeddings_pretrained_models} for word discrimination, is shown in Fig.~\ref{fig:meanpooling_results}.
The graphs show that choosing the correct layer at which feature meanpooling is applied leads to improvements over the DTW baseline for all three languages considered.
It seems that the central layers of these architectures encode the information most useful for AWEs~\cite{sanabria2023analyzing_embeddings_pretrained_models,Pasad2021LayerWiseAO}.

All four pre-trained models include English in their training sets.
Since W2V2 is monolingual, it is not surprising that it achieves the largest improvement for
English.
For the two under-resourced languages, which are the focus of this work, we see that MHB provides a clear improvement over the DTW baseline in both cases. {In particular, features from layer 8 yield the best overall performance.}
Since Luganda is present in the MHB  training corpus, this might be expected.
However, the results for the unseen language~(Bambara) indicate that the smaller MHB model trained on less data outperforms XLS-R.

\subsection{Keyword spotting using multilingual AWE models}
\label{sec:AWE_Model_Section}

Table~\ref{tbl:multilingual_AWE_dev_results} reports the KWS performance different embedding approaches on the development set.
Firstly, we observe that in addition to being less computationally efficient, AWEs obtained by subsampling are outperformed by meanpooling. 
All three multilingual AWE models outperform the 
methods derived directly from self-supervised features for our evaluation languages.
Among these, the two contrastive models outperform the CAE-RNN, which was previously applied to KWS on radio broadcasts, especially for Luganda.
The five different AWE approaches all improve on the DTW baseline. The greatest increases are observed for the two contrastive models, with absolute MAP improvements of 47\% and 34\% for Luganda and Bambara respectively.

The performance of the two contrastive models on the test set for our evaluation languages is shown in Table \ref{tbl:multilingual_test_results}. ContrastiveTransformer, our proposed model, is seen to outperform the ContrastiveRNN model on
all metrics for the test set.
{The increase in P@N indicates that the transformer model more consistently retrieves utterances containing the keyword.}

In contrast to the the two under-resourced languages, we note that for English, the multilingual AWE models do not exhibit as large an improvement over the pre-trained mHuBERT approaches.
We believe that this is because the mHuBERT training set includes a large amount of English.
These results nonetheless demonstrate that contrastive learning using features provided by large self-supervised models trained on well-resourced languages can lead to consistent improvements even in unseen languages when applied to downstream tasks such as keyword spotting.

\begin{table}[!t]
    \centering
    \scriptsize
    \caption{Test set keyword spotting results achieved by the two top-performing systems in Table~\ref{tbl:multilingual_AWE_dev_results}.}
    \vspace*{-2mm}
    \begin{tabularx}{\linewidth}{@{\extracolsep{-4pt}}lcccccc}
    \toprule
    & \multicolumn{3}{c}{Luganda} & \multicolumn{3}{c}{Bambara}  \\
    \cmidrule(lr){2-4} \cmidrule(lr){5-7}
    & MAP & P@10 & P@N & MAP & P@10 & P@N \\ 
    \midrule
        DTW with BNF (baseline) &   24.3 & 40.8 & 26.3 & 42.4 & 59.1 & 42.3 \\ 
        
        ContrastiveRNN & 60.6 & 75.0 & 60.8 & 68.9 & 81.3 & 67.6\\
        
        ContrastiveTransformer &  \textbf{65.3} & \textbf{75.8} & \textbf{63.4} & \textbf{69.9} & \textbf{82.2} & \textbf{68.1} \\
        
        \bottomrule
    \end{tabularx}
    \label{tbl:multilingual_test_results}
    \vspace{-5mm}
\end{table}

\subsection{Analysis of training language combinations}
\label{sec:AWE_Languages_results}

For rapid model deployment in low-resource languages, assessing the effectiveness of available training languages is impractical. Table~\ref{tbl:contrastiveRNN_MAP_comparison} shows KWS performance for different training language combinations. 
The Swahili monolingual model performs best for both Bambara and the related Luganda.
Among the multilingual models, the best Bambara performance is achieved using two diverse languages (sw+afr), whereas adding unrelated Southern Bantu languages (xho+sot) hurts performance.
We conclude that AWE models for KWS benefit from (1) training on related languages when these are available (e.g., Swahili for Luganda) and (2) prioritising using distinct languages over multiple closely related languages when none related to the target language are available.



\begin{table}[!h]
    \centering
    \scriptsize
    \vspace*{-2mm}
    \caption{MAP scores on the development set for different language combinations using the ContrastiveTransformer.}
    \vspace*{-2mm}
    \begin{tabularx}{\linewidth}{l>{\centering\arraybackslash}X>{\centering\arraybackslash}X} 
    \toprule
    Language combination (100K pairs per language) & Luganda & Bambara \\ 
    \midrule
    sw & 68.09 & 72.97 \\
    afr & 54.21 & 72.24 \\
    xho & 62.15 & 69.89 \\
    sw + afr & 67.59 & \textbf{75.02} \\ 
    sw + xho & 64.90 & 73.25 \\
    afr + sw + sot + xho (from Table \ref{tbl:multilingual_AWE_dev_results}) & \textbf{69.91} & 74.53 \\
    \bottomrule
    \end{tabularx}
    \label{tbl:contrastiveRNN_MAP_comparison}
    \vspace{-5mm}
\end{table}

\section{Conclusion}

We propose a new approach to very low-resource keyword spotting (KWS), the ContrastiveTransformer, which leverages acoustic representations computed by large pre-trained architectures and performs subsequent contrastive training of acoustic word embeddings (AWEs) using well-resourced languages.
The AWEs produced by this transformer are then used to encode speech in the target language.
We compare this approach with two other recent acoustic word embedding approaches, which also incorporate features obtained from large self-supervised models.
As further baselines, we consider direct meanpooling and subsampling of the features produced by the large pre-trained models, as well as a classical query-by-example system based on dynamic time warping (DTW).
While meanpooling and subsampling outperform the DTW baseline, the addition of contrastive training is shown to further improve KWS performance for Bambara and Luganda, two under-resourced African languages.
The ContrastiveTransformer architecture proposed in this paper achieved the largest performance improvements across all three considered languages.

\section{Acknowledgements}

We thank NVIDIA for donating GPU equipment, Ewald van der Westhuizen for the NCHLT alignments and Herman Kamper for valuable discussions. We also thank the DW Ackermann Bursary Fund and Telkom South Africa for support.

\bibliographystyle{IEEEtran}
\bibliography{refs}

\end{document}